# Search for anomalies in the $\nu_e$ appearance from a $\nu_\mu$ beam.

M. Antonello[1], B. Baibussinov[2], P. Benetti[3], F. Boffelli[3], A. Bubak[4], E. Calligarich[3],
N. Canci[1], S. Centro[2], A. Cesana[5], K. Cieslik[6], D. B. Cline[7], A.G. Cocco[8],
A. Dabrowska[6], D. Dequal[2], A. Dermenev[9], R. Dolfini[3], A. Falcone[3], C. Farnese[2],
A. Fava[2], A. Ferrari[10], G. Fiorillo[8], D. Gibin[2], S. Gninenko[9], A. Guglielmi[2],
M. Haranczyk[6], J. Holeczek[4], M. Kirsanov[9], J. Kisiel[4], I. Kochanek[4], J. Lagoda[11],
S. Mania[4], A. Menegolli[3], G. Meng[2], C. Montanari[3], S. Otwinowski[7], P. Picchi[12],
F. Pietropaolo[2], P. Plonski[13], A. Rappoldi[3], G.L. Raselli[3], M. Rossella[3],
C. Rubbia[1,10,14], P. Sala[5], A. Scaramelli[5], E. Segreto[1], F. Sergiampietri[16], D. Stefan[1],
R. Sulej[11,1], M. Szarska[6], M. Terrani[5], M. Torti[3], F. Varanini[2], S. Ventura[2],
C. Vignoli[1], H. Wang[7], X. Yang[7], A. Zalewska[6], A. Zani[3], K. Zaremba[13]

(ICARUS Collaboration)

[1] *INFN - Laboratori Nazionali del Gran Sasso, Assergi, Italy*
[2] *Dipartimento di Fisica e Astronomia Università di Padova and INFN, Padova, Italy*
[3] *Dipartimento di Fisica Università di Pavia and INFN, Pavia, Italy*
[4] *Institute of Theoretical Physics,Wroclaw University, Wroclaw, Poland*
[5] *Politecnico di Milano and INFN, Milano, Italy*
[6] *H. Niewodniczanski Institute of Nuclear Physics, Polish Academy of Science, Krakow, Poland*
[7] *Department of Physics and Astronomy, UCLA, Los Angeles, USA*
[8] *Dipartimento di Scienze Fisiche Università Federico II di Napoli and INFN, Napoli, Italy*
[9] *INR RAS, Moscow, Russia*
[10] *CERN, Geneva, Switzerland*
[11] *National Centre for Nuclear Research, Otwock/Swierk, Poland*
[12] *Institute of Physics, University of Silesia, Katowice, Poland*
[13] *INFN Laboratori Nazionali di Frascati, Frascati, Italy*
[14] *GSSI, L'Aquila, Italy*
[15] *Institute of Radioelectronics, Warsaw University of Technology, Warsaw, Poland*
[16] *INFN, Pisa, Italy*

## ABSTRACT

We report an updated result from the ICARUS experiment on the search for
$\nu_\mu \rightarrow \nu_e$ anomalies with the CNGS beam, produced at CERN with an average energy
of 20 GeV and travelling 730 km to the Gran Sasso Laboratory. The present analysis
is based on a total sample of 1995 events of CNGS neutrino interactions, which
corresponds to an almost doubled sample of the previously published result. Four
clear $\nu_e$ events have been visually identified over the full sample, compared with an
expectation of 6.4 ± 0.9 events from conventional sources. The result is compatible
with the absence of additional anomalous contributions. At 90% and 99% confidence
levels, the limits to possible oscillated events are 3.7 and 8.3 respectively. The
corresponding limit to oscillation probability becomes consequently $3.4 \times 10^{-3}$ and 7.6
$\times 10^{-3}$ respectively. The present result confirms, with an improved sensitivity, the
early result already published by the ICARUS Collaboration.

PACS. 14.60.ST Non-standard-model neutrino, right-handed neutrinos, etc. - 29.40.-n

ICARUS [1,2] is a large mass LAr-TPC imaging detector located at the Gran Sasso underground laboratory, 730 km away from the CERN neutrino source. It has an instrumented mass in excess of 476 ton of liquid Argon (LAr) and provides a completely uniform imaging of neutrino events with accuracy, density and interaction lengths similar to the ones of a heavy Freon conventional bubble chamber. This innovative detection technique allows observing the actual "3D-image" of each charged track with a resolution of few $mm^3$.

The CNGS neutrino facility [3] provides an almost pure $\nu_\mu$ beam peaked in the range $10 \leq E_\nu \leq 30$ GeV, with an electron component of less than 1% [4]. From October 2010 to December 2012, we have collected a total of neutrino data corresponding to $8.6 \times 10^{19}$ POT (400 GeV protons on target) and with the excellent recording efficiency exceeding 93%.

The LSND experiment [5] at the LANSCE Los Alamos accelerator and the MiniBooNE experiment [6] at the FNAL-Booster have previously reported significant evidence for an anomalous excess of $\nu_\mu \to \nu_e$ and $\bar{\nu}_\mu \to \nu_e$ at $L/E_\nu \approx 0.5 \div 1.0$ m/MeV where L is the distance from the target and $E_\nu$ is the neutrino energy. These results may imply the presence of an additional mass-squared difference somewhere within a wide interval $\Delta m^2_{new} \approx 0.01$ to $1.0$ eV$^2$ and with a corresponding associated value of $\sin^2(2\theta_{new})$, largely in excess of the predictions of the Standard Model and three neutrino mixing. Additional $\nu_e$ or $\bar{\nu}_e$ disappearance anomalies have been observed at similar $\Delta m^2_{new}$ values in (a) nearby nuclear reactors [7] and (b) Mega-Curie k-capture calibration sources [8,9].

In our case, such anomalies due to the $\nu_e$ appearance in a $\nu_\mu$ beam will be observed at much larger values of $L/E_\nu$ , centered around $L/E_\nu \approx 36.5$ m/MeV. These hypothetical anomalies will therefore produce very fast oscillations as a function of E, averaging over the observed spectrum to $\sin^2(1.27 \ \Delta m^2_{new} L/E.) \approx 1/2$ and $\left\langle P\left(\nu_\mu \to \nu_e\right)\right\rangle = 1/2 \sin^2(2\theta_{new})$.

A previous search for such anomalies in the CNGS neutrino beam has been recently published by the ICARUS Collaboration [4], based on 1091 neutrino events within the sensitive LAr volume and $3.3 \times 10^{19}$ POT. We have shown that there was a possible agreement of all published experimental results only for a narrow surviving region centred around $(\Delta m^2, \sin^2(2\theta))_{new} = (0.5$ eV$^2$, $0.005)$. In this paper we present an additional event sample of 904 neutrino events, bringing the total to 1995 events and $6 \times 10^{19}$ POT.

As described in more detail in Ref. [4] the neutrino interaction vertex and 2D projections of tracks and showers are identified visually. The event reconstruction is based on the signals recorded by the three TPC wire planes [2, 10] at angles 60° apart. After hit finding and fitting, the energy deposition is computed in the charge collecting view. A correction is introduced based on the (small) electron signal attenuation due to the drift distance directly measured with the help of cosmic ray muons. The high density of sampling - corresponding to ~ 2% of a radiation length - and the remarkable signal/noise ratio of about 10/1 allow to measure the specific ionisation of each wire. It is also possible to perform precise calorimetry and particle identification for stopping particles [10] and obtain a powerful electron/$\gamma$ separation [4]. The total visible energy of the events has been determined from the total charge collected by the TPC wires, corrected for the electronic response [2] and for the dE/dx recombination of the signals in LAr [11].

A sophisticated MonteCarlo simulation package dedicated to the ICARUS T600 detector has been developed [4]. It includes a neutrino event generator [12] accounting for quasi-elastic, resonant and deep inelastic interactions and describes the effects of Fermi motion, Pauli blocking and other initial and final state effects like, for instance, re-interactions of the reaction products inside the target nucleus [13]. The products of the neutrino interaction are then transported, with a detailed simulation of the energy losses and electromagnetic and hadronic interactions, including recombination effects [11]. In order to realistically reproduce the actual wire signals as recorded in the events, the response of the electronics and the noise patterns estimated from the data have been carefully simulated.

Both local energy deposition by muon, proton and pion tracks and global calorimetric reconstruction for $\nu-$CC interactions confirm that the detector response is reproduced to better than 2.5%, and the effective noise level is correctly simulated [4]. An ongoing study on low energy showers from isolated secondary $\pi^0$'s confirms that MonteCarlo reproduces experimental data for the ionisation at the beginning of the e.m. showers, a key tool for the powerful electron–photon discrimination [4]. We observe a general agreement between expectations of the MonteCarlo and the actually observed number of events.

Following the previous analysis [4], interaction vertices at a distance less than 5 cm from each side of the active volume of the TPC or less than 50 cm from its downstream walls have been discarded from the recorded sample. The "electron neutrino signature" has been defined [4] requiring:
- interaction vertex located inside the previously defined fiducial volume;
- event energy E < 30 GeV, in order to reduce the beam $\nu_e$ background;
- a primary charged track starting directly from the vertex, fully consistent over at least 8 wire hits with a minimum ionising relativistic particle (i.e. dE/dx < 3.1 MeV/cm on average after removal of visible delta rays) and subsequently building up into a shower;
- the electron candidate track has to be spatially separated from other ionising tracks within 150 mrad in the immediate proximity of the vertex in at least one of the two transverse views (± 60°), except for short proton like recoils due to nuclear interactions.

The expected number of $\nu_e$ events due to conventional sources in the energy range and fiducial volume are:
- 5.7 ± 0.8 events due to the estimated $\nu_e$ beam contamination;
- 2.3 ± 0.5 $\nu_e$ events due to the $\nu_\mu \rightarrow \nu_e$ oscillations from $\sin^2(\theta_{13}) = 0.0242 \pm 0.0026$;
- 1.3 ± 0.1 $\nu_\tau$ with $\tau \rightarrow$ e events from the three neutrino mixing standard model predictions,

giving a total of 9.3 ± 0.9 expected events, where the errors represent the uncertainty on the NC and CC contributions.

The selection efficiency for the search of a $\nu_e$ anomaly has been previously estimated as $\eta = 0.74 \pm 0.05$ [4] in the selected energy region. For the intrinsic $\nu_e$ contamination the slightly lower value 0.65 ± 0.06 has been estimated since its spectrum is harder than the one of the expected anomalies, based on a sample of 300 simulated events. The contribution from misidentified $\nu_\tau$CC and $\nu$NC interactions is negligible, as discussed in [4]. The predicted visible background is then 6.4 ± 0.9 (syst. error only) events. A thorough discussion on the estimate of the systematic uncertainties on the predicted number of $\nu_e$ events was already presented in the previous ICARUS paper on the search for the LSND anomaly [4].

In the newly added sample we have found two additional electron events that bring to four the total observed number of events. This is compatible with the expectation of 6.4 ± 0.9 due to conventional sources: the probability to observe a statistical under-fluctuaction resulting in four or less $\nu_e$ events is 25%.

The first new event, shown in Figure 1, has a total energy of ~27 GeV and an electron of 6.3 ± 1.5 GeV, taking into account the partially escaping fraction of the e.m. showers. The electron is clearly separated from the other tracks after 1 cm from the main vertex. The progressive evolution of the electron from the single ionising particle to an electromagnetic shower is clearly visible in the plot of dE/dx along the individual wires in Figure 1.

The second new event, shown in Figure 2, has a total energy of ~14 GeV and an electron of 6.4 ± 0.3 GeV. The corresponding three-dimensional reconstruction of the event is also shown.

In both events the single electron shower in the transverse plane is opposite to the remaining of the event, with the electron transverse momentum of 3.5 ± 0.9 GeV/c and 1.2 ± 0.2 GeV/c respectively.

Our previously published result [4] is therefore extended with an almost doubled event statistics. At statistical confidence levels of 90% and 99% and taking into account the revised detection efficiency $\eta$, the limits are respectively 3.7 and 8.3 events [14]. The corresponding new limits on the oscillation probability are $\left\langle P\left(\nu_\mu \to \nu_e\right)\right\rangle \leq 3.4 \times 10^{-3}$ and $\left\langle P\left(\nu_\mu \to \nu_e\right)\right\rangle \leq 7.6 \times 10^{-3}$ respectively.

The new exclusion area of the ICARUS experiment referred to neutrino-like events is shown in Figure 3 in terms of the two dimensional plot of $\sin^2(2\theta_{new})$ and $\Delta m^2_{new}$. In the interval $\Delta m^2_{new} \approx 0.1$ to >10 eV$^2$ the exclusion area is independent of $\Delta m^2_{new}$ with $\sin^2(2\theta_{new}) = 2.0\left\langle P\left(\nu_\mu \to \nu_e\right)\right\rangle$. In the $\Delta m^2_{new}$ interval from 0.1 to $\approx 0.01$ eV$^2$, the oscillation is progressively growing and averages to about the above value of twice $\left\langle P\left(\nu_\mu \to \nu_e\right)\right\rangle$. For even lower values of $\Delta m^2_{new}$, the longer baseline strongly enhances the oscillation probability with respect to the one of the previous short baseline experiments.

The LSND result [5] was based on anti-neutrino events. A small ~2% anti-neutrino event contamination is also present in the CNGS beam as experimentally observed [15]. According to a detailed neutrino beam calculation, the $\bar{\nu}_\mu$ CC event rate is (1.2 ± 0.25) % for $E_\nu < 30$ GeV, where a 20 % uncertainty has been conservatively assumed. In the limiting case in which the whole effect is due to $\bar{\nu}_\mu \to \bar{\nu}_e$, the absence of an anomalous signal gives a limit of 4.2 events at 90% CL. The corresponding limit on the oscillation probability is $\left\langle P(\bar{\nu}_\mu \to \bar{\nu}_e)\right\rangle \leq 0.32$. The resulting (small) exclusion area is shown in Figure 3.

As shown in Figure 3, a major fraction of the initial two dimensional plot [$\Delta m^2$, $\sin^2(2\theta)$]$_{new}$ of the main published experiments sensitive to the $\nu_\mu \to \nu_e$ anomaly [5,6,16,17,18] is now excluded by the present result.

The MiniBooNE [6] experiment has recorded both antineutrino and neutrino data. The LSND result relates to the antineutrino signal and it is statistically significant only for $E_\nu/L > 1$ MeV/m, corresponding in the MiniBooNE conditions to $E_\nu^{QE} > 475$ MeV. In this energy region, a significant LSND-like effect is still observed for antineutrino while a much weaker evidence, compatible with the absence of a signal is apparent in the neutrino data. This incompatibility has been explained in Ref. [6] as caused by a number of possible reasons, like expanded oscillation models with several sterile neutrinos, CP violating effects and so on or by unpredicted systematic uncertainties and backgrounds. Therefore there is tension and the compatibility between the MiniBooNE antineutrino and neutrino data is low, at least in a simple two-neutrino oscillation model.

In the MiniBooNE region $200 < E_\nu^{QE} < 475$ MeV — below the sensitive $E_\nu/L$ region of LSND — a *new effect* and a significant additional anomaly has been reported [6] both for neutrino and antineutrino data. The neutrino result may be compared with the present experiment.

The present experiment has observed electron events at much larger values of $L/E_\nu$, centered around $L/E_\nu \approx 36.5$ m/MeV. In order to compare the results with LSND and MiniBooNE, the values of the oscillation probability have to be projected to lower values of $L/E_\nu$. The two-neutrino model $P = \sin^2(2\theta)\sin^2(1.27\Delta m_{41}^2 L/E_\nu)$ has been used to calculate the $\nu_\mu \to \nu_e$ oscillation probability as a function of the neutrino energy $E_\nu$ from the observed number of excess events/MeV of Figure 2 in Ref. [6]. The conversion has been extracted directly from the above graph of Ref. [6], converting the ratio of the excess events/MeV to the oscillation probability using their (also plotted) example of the two neutrino model case with $\sin^2(2\theta) = 0.2$ and $\Delta m_{41}^2 = 0.1$ eV$^2$ from Figure 2 of [6].

The result is shown in Figure 4. There is tension between the limit $\sin^2(2\theta_{new}) < 6.8\times10^{-3}$ at 90% CL and $< 1.52\times10^{-2}$ at 99% CL of the present experiment and the neutrino lowest energy points of MiniBooNE with $200 < E_\nu^{QE} < 475$ MeV, suggesting an instrumental or otherwise unexplained nature of the low energy signal reported by Ref. [6]. Recently a similar search performed at the same CNGS beam by the OPERA experiment has confirmed our finding and the absence of anomalous oscillations with an independent limit $\sin^2(2\theta_{new}) < 7.2\times10^{-3}$ [19].

As a conclusion, the LSND anomaly appears to be still alive and further experimental efforts are required to prove the possible existence of sterile neutrinos. The recently proposed ICARUS/NESSiE experiment at the CERN-SPS neutrino beam [20], based on two identical LAr-TPC detectors, complemented with magnetized muon spectrometers and placed at two different distances from proton target, has been designed to definitely settle the origin of these ν-related anomalies.

### ACKNOWLEDGMENTS

The ICARUS Collaboration acknowledges the fundamental contribution to the construction and operation of the experiment given by INFN and, in particular, by the LNGS Laboratory and its Director. The Polish groups acknowledge the support of the Ministry of Science and Higher Education, and of National Science Centre, Poland. Finally, we thank CERN, in particular the CNGS staff, for the successful operation of the neutrino beam facility.

**REFERENCES**

[1] C. Rubbia et al. [ICARUS Coll.], JINST 6 P07011 (2011) and references therein.

[2] S. Amerio et al. [ICARUS Coll.], Nucl. Instrum. and Methods Phys. Res., Sect. A 527, 329 (2004).

[3] G. Acquistapace et al. [CNGS Coll.] CERN 98-02, INFN/AE-89-05 (1998); R. Bailey et al. CERN-SL/99-034 (DI), INFN/AE- 99/05 Addendum (1999); E. Gschwendtner et al., CERN-ATS-2010-153 (2010).

[4] M. Antonello et al. [ICARUS Coll.], Eur. Phys. J. C, 73:2345 (2013).

[5] A. Aguilar et al. [LSND Coll.], Phys. Rev. D 64, 112007 (2001).

[6] A. A. Aguilar-Arevalo et al. [MiniBooNE Coll.],PRL 110, 161801 (2013) and references therein.

[7] G. Mention et al., Phys.Rev. D83 (2011) 073006 and references therein.

[8] J. N. Abdurashitov et al. [SAGE Coll.], Phys. Rev. C 80, 015807 (2009).

[9] F. Kaether, W. Hampel, G. Heusser, J. Kiko, and T. Kirsten [GALLEX], Phys. Lett. B 685, 47 (2010) and references therein.

[10] M. Antonello et al. [ICARUS Coll.], Adv. High Energy Phys.  ,260820 (2013).

[11] S. Amoruso et al. [ICARUS Coll.], Nucl. Instrum. Methods Phys. Res., Sect. A 523, 275 (2004).

[12] G. Battistoni et al. [FLUKA Coll.], in Proceedings of the 12th International Conference on nuclear reaction mechanisms, Varenna, Italy, June 15-19 (2009), p.307.

[13] F. Arneodo et al. [ICARUS and Milano Coll.], Phys. Rev. D 74, 112001 (2006).

[14] G.J. Feldman and R. D. Cousins, Phys. Rev. D 57, 3873 (1998).

[15] N. Agafonova et al. [OPERA Coll.], New J. Phys. 13 (2011), 053051.

[16] B. Armbruster et al. [KARMEN Coll.], Phys. Rev. D 65, 112001 (2002).

[17] P. Astier et al. [NOMAD Coll.], Phys. Lett. B 570 19 (2003).

[18] S. Avvakumov et al. [NuTeV Coll.], Phys. Rev. Lett. 89 (2002) 011804.

[19] N. Agafonova et al. [OPERA Coll.], JHEP 1307 (2013) 004.

[20] M. Antonello et al. [ICARUS/NESSiE Coll.], CERN-SPSC-2012-010 and SPSC-P-347 (2012).

**FIGURES**

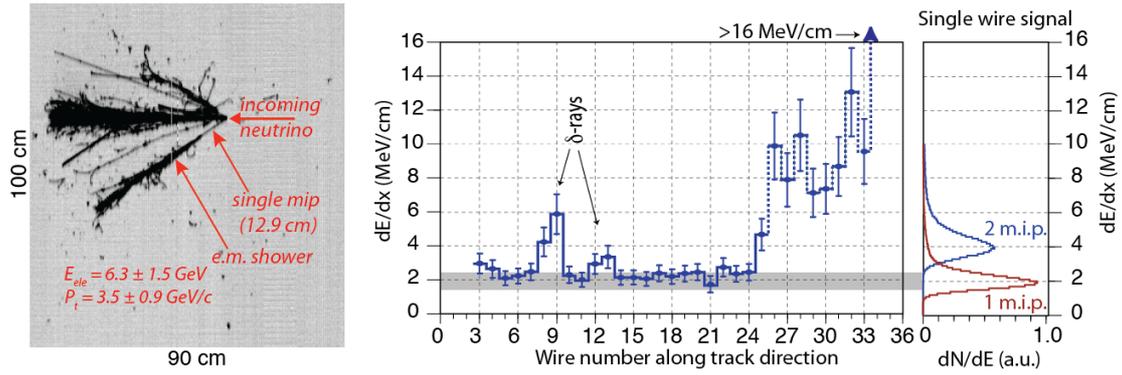

**Figure 1.** Experimental pictures of the first of the two events with a clear electron signature found in the additional sample of 904 neutrino interactions. The evolution of the actual dE/dx from a single track to an e.m. shower for the electron shower is shown along the individual wires. The event has a total energy of ~27 GeV and an electron of 6.3 ± 1.5 GeV with a transverse momentum of 3.5 ± 0.9 GeV/c.

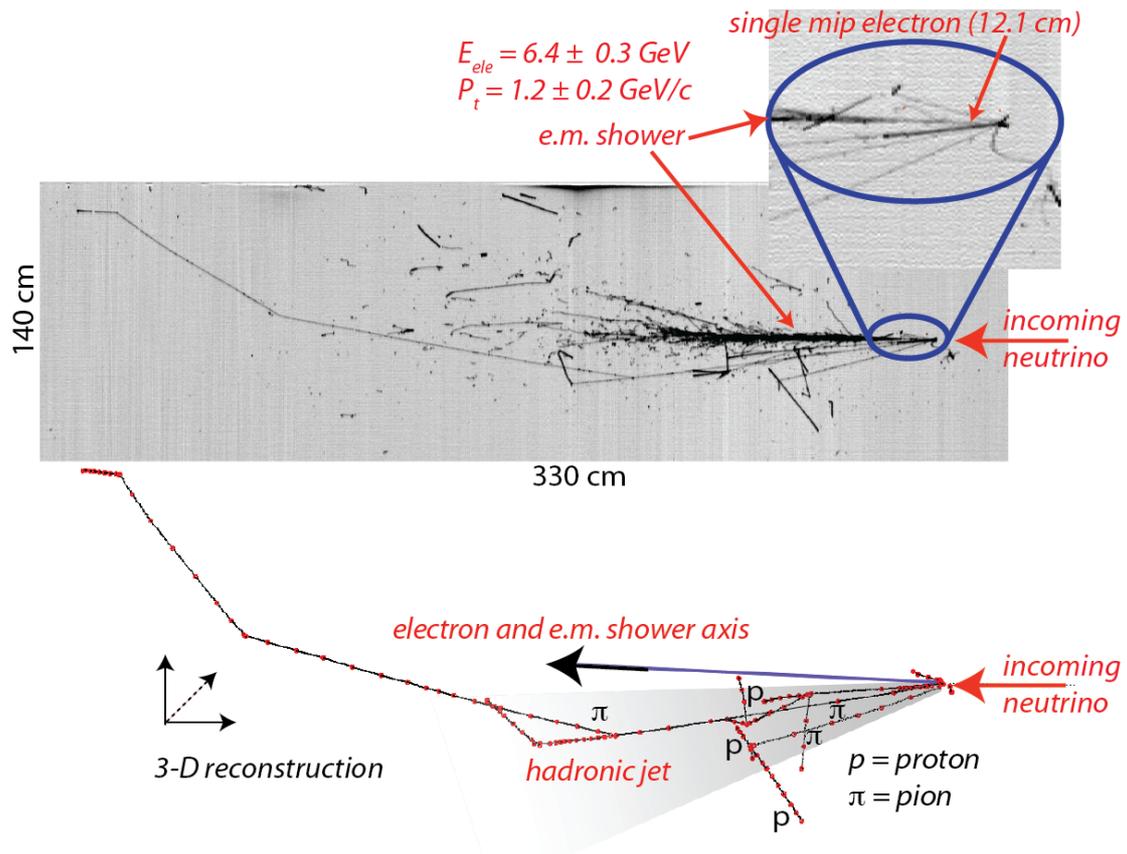

**Figure 2.** Second $\nu_e$ event. It has a total energy of ~14 GeV and an electron of 6.4 ± 0.3 GeV with transverse momentum of 1.2 ± 0.2 GeV/c. The 3D reconstruction of primary particles in the event is also shown (red dots correspond to vertices of polygonal fit [10]).

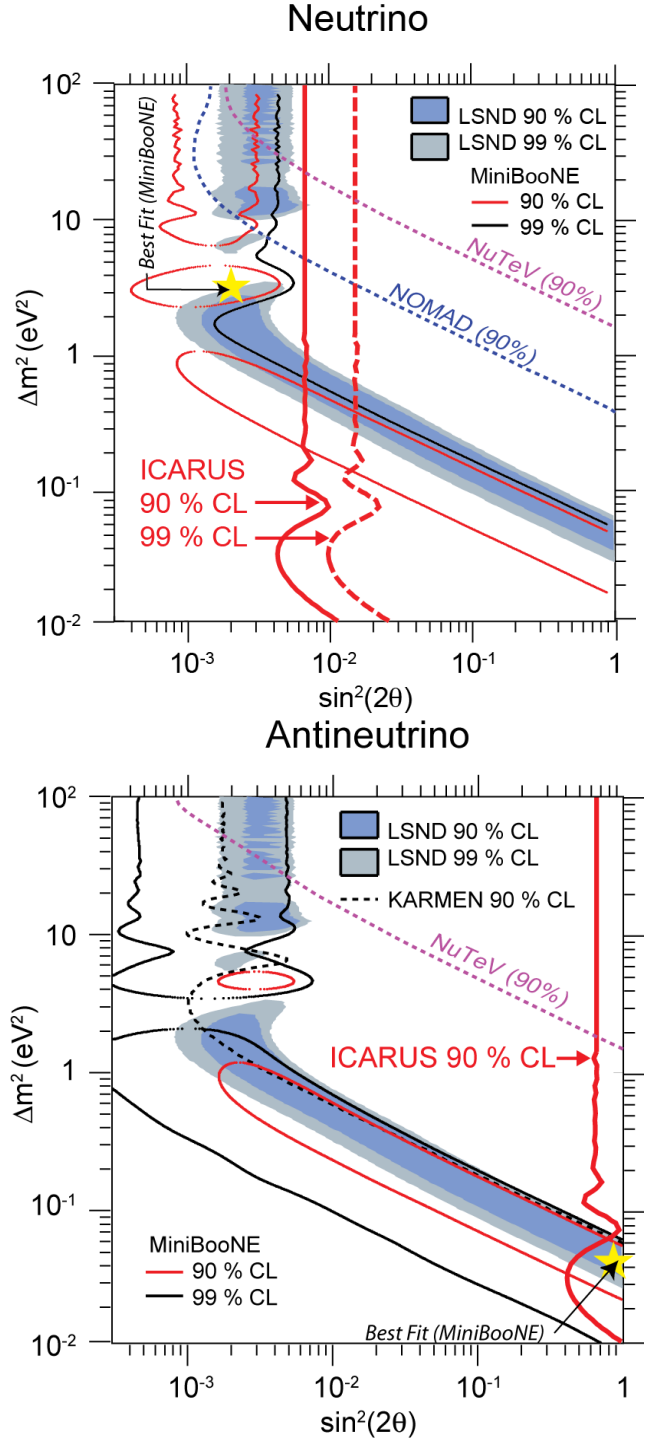

**Figure 3.** Neutrino (top) and antineutrino (bottom) with $\Delta m^2$ as a function of $\sin^2(2\theta_{new})$ for the main experiments sensitive to the $\bar{\nu}_\mu \rightarrow \bar{\nu}_e$ and $\nu_\mu \rightarrow \nu_e$ anomalies [5,6,16,17,18] and for the present result (continuous red lines). The yellow stars mark the best fit points of MiniBooNE [6]. The ICARUS limits on the oscillation probability for $\nu_\mu \rightarrow \nu_e$ are $\left\langle P\left(\nu_\mu \rightarrow \nu_e\right)\right\rangle \leq 3.4 \times 10^{-3}$ and $\left\langle P\left(\nu_\mu \rightarrow \nu_e\right)\right\rangle \leq 7.6 \times 10^{-3}$ at 90% and 99% CL, corresponding to $\sin^2(2\theta_{new}) \leq 6.8 \times 10^{-3}$ and $\sin^2(2\theta_{new}) \leq 1.5 \times 10^{-2}$ respectively. The ICARUS limit on the $\bar{\nu}_\mu \rightarrow \bar{\nu}_e$ oscillation probability is $\left\langle P\left(\bar{\nu}_\mu \rightarrow \bar{\nu}_e\right)\right\rangle \leq 0.32$ at 90% CL, corresponding to $\sin^2(2\theta_{new}) \leq 0.64$.

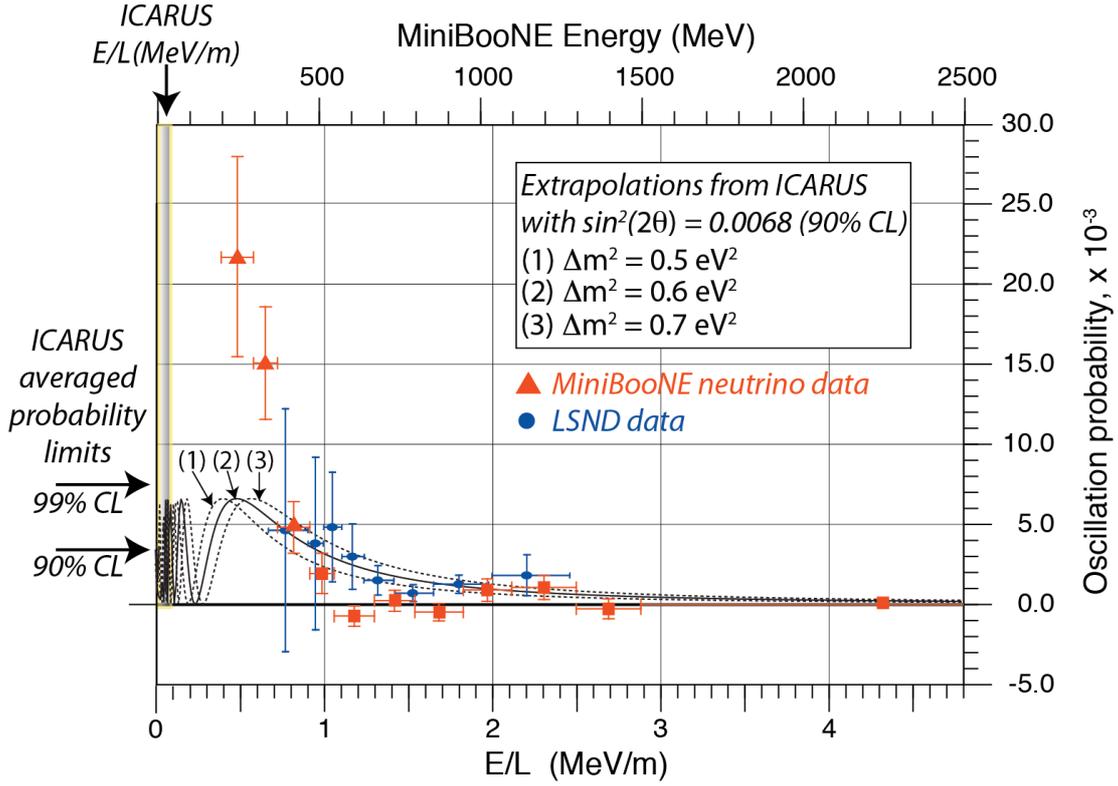

**Figure 4.** Oscillation probability limits coming from the present experiment compared with corresponding data from neutrinos in MiniBooNE [6] as a function of $E_\nu / L$. Figure 2 in Ref. [6] has been used in order to convert the observed number of excess events/MeV to their corresponding oscillation probabilities. In order to perform the conversion, the values $\sin^2(2\theta) = 0.2$ and $\Delta m^2_{41} = 0.1$ eV$^2$ from Figure 2 of Ref. [6] have been used. The resulting oscillation probability distribution for neutrino and for $E_\nu > 475$ MeV (square red points) appears incompatible with the antineutrino LNSD effect. In the $200 < E_\nu^{QE} < 475$ MeV region (triangular red points) — below the sensitive $E_\nu/L$ region of LSND — the new MiniBooNE effect is widely incompatible with the averaged upper probability limits to anomalies from the present paper and from OPERA [19] in their $E_\nu/L$ regions. An extrapolation from ICARUS (black curves marked as 1, 2 and 3) to larger values of E/L for two-neutrino oscillation parameters simultaneously compatible with LSND, MiniBooNe and Karmen is also shown as guidance.